\batchmode
\makeatletter
\def\input@path{{/home/flatmax/flatmax/personal.work/research/HearingAid/}}
\makeatother
\documentclass[australian]{article}
\usepackage[T1]{fontenc}
\usepackage[latin9]{inputenc}
\usepackage{geometry}
\usepackage{amsmath}
\usepackage{graphicx}

\makeatletter
\@ifundefined{showcaptionsetup}{}{%
 \PassOptionsToPackage{caption=false}{subfig}}
\usepackage{subfig}
\makeatother

\usepackage{babel}
\begin{document}
\title{The Learning Prescription, A Neural Network Hearing Aid Core}
\author{Matthew R. Flax <matt.flax@flatmax.com>}
\maketitle
\begin{abstract}
The definition of a hearing aid core which is based on a prescription
neural network (such as NAL-NL2) is defined here. This hearing aid
core replaces a traditional compressor hearing aid core which mimics
the said hearing aid prescription. Whilst the replacement of the compressors
for a neural network may seem simple, the implications are vast in
terms of the ``learning prescription'' where the topology of the
neural network may be increased to make available more free parameters
and allow great personalisation of the hearing aid prescription.
\end{abstract}

\section{Introduction}

The NAL-NL2 hearing aid prescription introduced a neural network for
the prescription of hearing aid gain for the first time \cite{Flax09}
based on a desensitised speech intelligibility index (SII) designed
for NAL-NL2 \cite{Flax.ssi.09}. Concise descriptions of the NAL-NL2
hearing aid prescription are given \cite{keidser2010derivation,keidser2011nal}
which focus on the effects of the desensitised SII on gain optimisation,
however the said articles gloss over the importance of the introduction
of the neural network to hearing aid prescription, which overcame
significant hurdles of reliable prescriptions being dispensed by NAL-NL1.
The reason why arbitrary prescription is now far more accurate was
the ability for the NAL-NL2 neural network to successfully interpolate
between optimised prescriptions for people with unique and unseen
hearing loss profiles. Prior to the introduction of the neural network
in hearing aid prescription, hand crafted nonlinear equations were
used to try to match the infinite possible prescriptions which can't
all be optimised and thus certain patients would not receive optimal
hearing aid prescriptions.

This article takes the next logical step in hearing aid development
by defining for the first time the replacement of hearing aid compressors
by a personal prescription neural network. This article lays the foundation
for the future layering of neural network and other statistically
optimised systems to greatly improve hearing aid performance. With
the introduction of personal prescription neural networks this article
also introduces a robust method for further personalisation away from
speech intelligibility prescriptions and towards learning prescriptions.

\begin{figure}
\begin{centering}
\subfloat[A digital hearing aid core. A filter bank bands a signal which is
compressed, summed and output. The compressors are fitted as best
as possible to the gain prescription. The signal chain is nonlinear.\label{fig:A-digital-hearing}]%
{\begin{centering}
\includegraphics[width=0.65\linewidth]{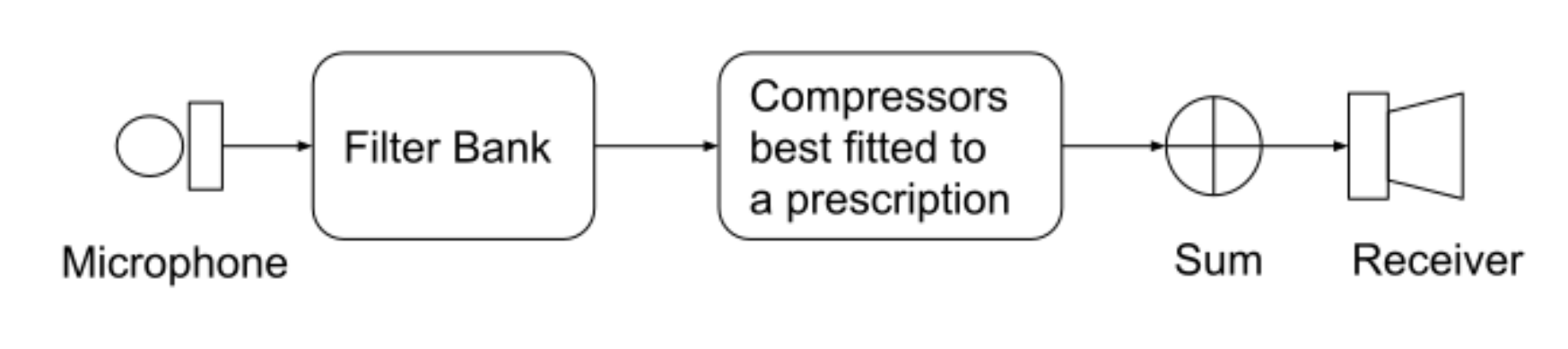}
\par\end{centering}
}\\
\subfloat[A neural network hearing aid core. The filter bank bands the signal
and a sound level meter (SLM) drives the prescription neural network.
The gains prescribed by the NN are applied to the signal which is
then summed and output. The signals in this core is block linear as
the gains are constant for each signal block. \label{fig:A-neural-network}]%
{\begin{centering}
\includegraphics[width=0.65\linewidth]{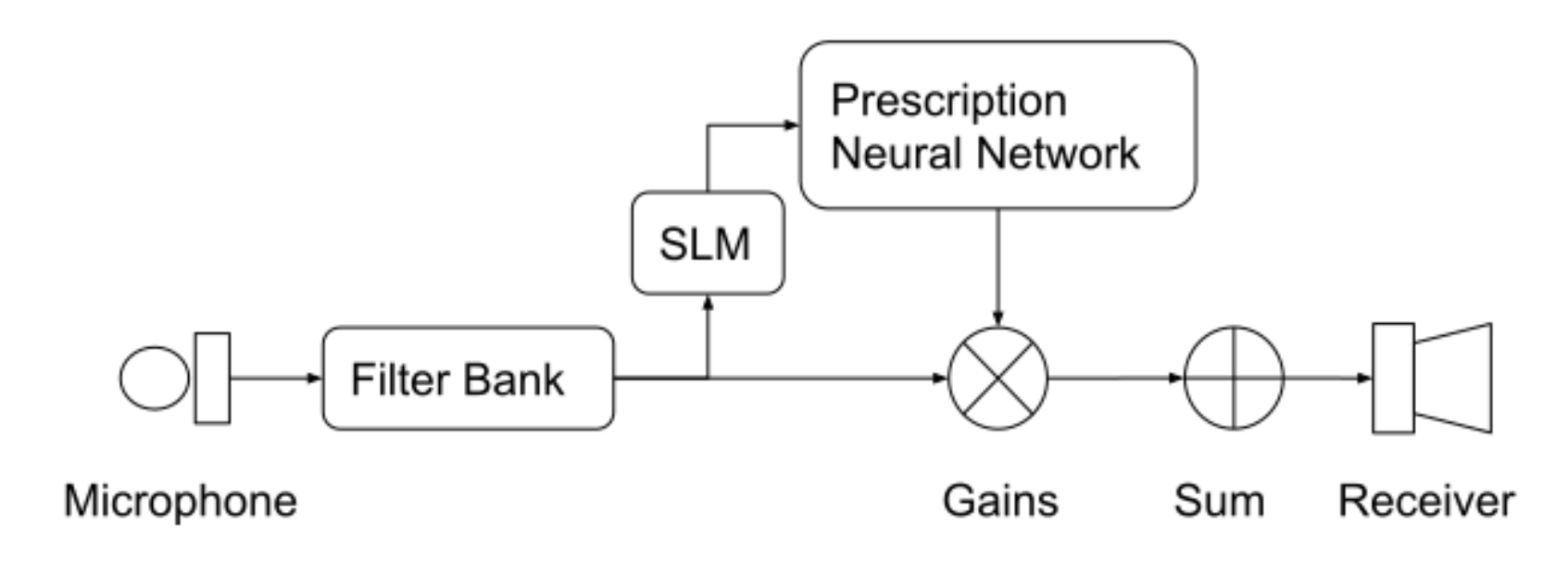}
\par\end{centering}
}
\par\end{centering}
\caption{Digital and neural network hearing aid cores.}
\end{figure}

A digital hearing aid core is shown in Figure \ref{fig:A-digital-hearing}
where a filter bank bands the signal and pre-fit compressors implement
the hearing aid prescription. The signal path is nonlinear as the
sound pressure level is constantly changing and the level estimation
in the compressors are constantly changing. This constantly changing
level estimation generates nonlinear gain application as the compressor's
operating point is slowly but constantly varying.

The hearing aid implemented with a prescription neural network core,
shown in Figure \ref{fig:A-neural-network} operates on a block of
N samples of audio signal. The sound level meter (SLM) presents signal
levels for the neural network to prescribe the block gain for each
band of the filter bank. The gains are applied to the banded signals
and summed then output to the receiver. As the gains are not varying
within a signal block, the signal chain is linear. Half window overlap
add techniques can be used to allow the audio blocks to vary smoothly
and allow the gains to vary without output discontinuity.

This article prescribes the implementation of a hearing aid with a
neural network core. Free software is also available which implements
the theory in this document. The first Section \ref{sec:Log-banded-filter}
implements a log banded filter bank centred around the prescription
frequencies ($f_{c}$). The duration of the audio in each filter is
roughly eight milliseconds and after overlap add the effective hearing
aid gains change at a rate of approximately four milliseconds. Rates
of gain change slower then three milliseconds are optimal for a prescription
algorithm such as NAL-NL2 \cite{Flax09} as the compression ratio
of the optimised prescription is not altered and thus the speech intelligibility
index is maximised. The overall latencies of the filters are half
the filter length as there is an overlap add framework. In operation,
the first half block can be output after half the filter length is
input/output and every subsequent half block is processed and output,
resulting in an overall latency of half the filter length which is
around 3 ms to 4 ms.

Subsequent sections \ref{sec:The-sound-level} and \ref{sec:Audio-amplification-and}
briefly address level metering and signal amplification. While the
last section leaves the prescription neural network as an open design
solution. The best neural network will start the user in a space which
is optimised for SII maximisation, but allow the user to train their
prescription to their own personal target. The gradual expansion of
the free parameters available to the neural network will allow for
the expansion in the complexity of gain prescription to the user's
taste.

\section{Log banded filter bank\label{sec:Log-banded-filter}}

The prescription algorithm outputs gains for the log centred frequencies
($f_{c}$ in Hz) over the M=6 bands from m=0 to m=M-1
\begin{align*}
f_{c}\left(m\right)= & 250\left(2^{m}\right)
\end{align*}

Zero phase brick wall band limited filters are generated\footnote{See the script gtkiostream/mFiles/ImpBandLim.m as a reference.}
where the zero phase filters ($h_{0,\,m}$) are specified in the Discrete
Fourier Domain ($H_{0,\,m}$) and transformed to the time domain using
the inverse Discrete Fourier Transform (DFT or $\mathcal{F}$)
\begin{align*}
H_{0,\,m}\left(f_{i}\left(m\right),\,f_{a}\left(m\right)\right) & =\left.1\right]_{f_{i}\le\left|f\right|\le f_{a}}\\
h_{0,\,m} & =\mathcal{F}^{-1}\left\{ H_{0,\,m}\right\} 
\end{align*}
where the minimum frequency ($f_{i}\left(m\right)$) and maximum frequency
($f_{a}\left(m\right)$) are specified per band m (see Equation \ref{eq:band-limits}).

These zero phase filters are circularly shifted by a constant group
delay of $\frac{N}{2}$ samples to give the linear phase band limited
filters ($h_{m}$)
\begin{align*}
h_{m}=h_{m,\,n} & =h_{m}[n]=h_{0,\,m}\left[\left(n+\frac{N}{2}\right)\,mod\,N-1\right]
\end{align*}

The specifiaction of the band limits (in Hz) are
\begin{align}
f_{a}\left(m\right)= & \begin{cases}
f_{c}\left(m\right)+f_{t} & m=0\\
\frac{3}{2}f_{c}\left(m\right) & m>0
\end{cases}\label{eq:band-limits}\\
f_{i}\left(m\right)= & \begin{cases}
20 & m=0\\
f_{a}\left(m-1\right) & m>0
\end{cases}\nonumber 
\end{align}
where $f_{t}$ is the frequency stepping between Fourier bins or the
DFT resolution, which is kept to a maximum value
\[
f_{t}=\frac{f_{c}\left(0\right)}{2}
\]
and this defines the number of samples (N) in the filter given a sample
rate of $f_{s}$ Hz
\[
N=\frac{f_{s}}{f_{t}}
\]

\begin{figure}
\begin{centering}
\includegraphics[width=0.65\linewidth]{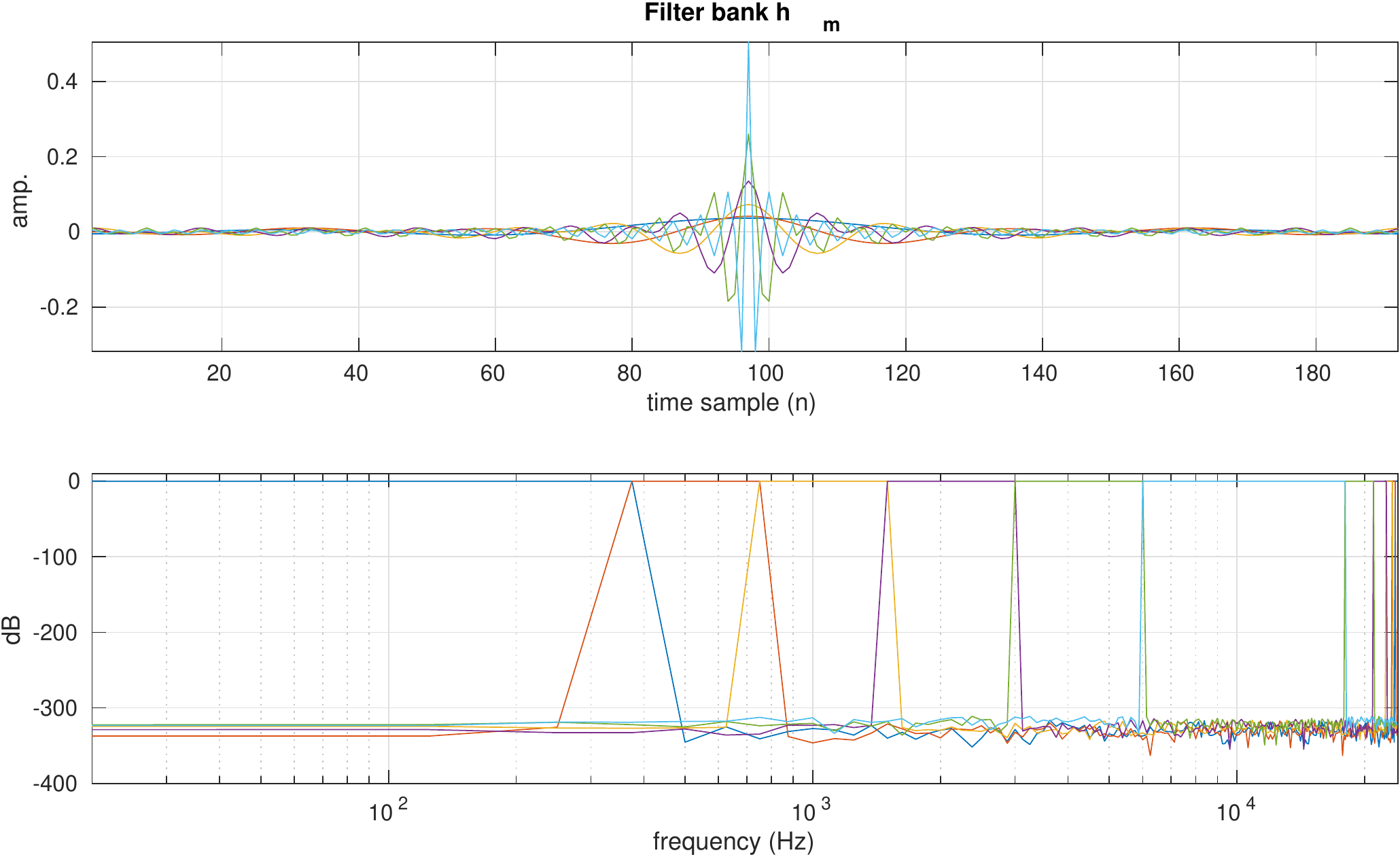}
\par\end{centering}
\caption{The hearing aid filter bank $h_{m}$.\label{fig:hearing-aid-filters}}
\end{figure}

An example filter bank with a sample rate of $f_{s}=24\,kHz$ is implemented
in the script LogFilterBankTest.m and is shown in Figure \ref{fig:hearing-aid-filters}.

\section{The sound level meter\label{sec:The-sound-level}}

The SLM estimates the dB SPL level of the signal ($s$) for each band
($l_{m}$)
\[
l_{m}=20\,log10\left(\sum_{n=0}^{N-1}h_{m}*s+l_{t,m}\right)+l_{d,m}
\]
where $*$ represents the convolution operator and the three scaling
variables are defined as; $l_{t}$ is a time domain DC offset which
may be necessary in some systems. $l_{d}$ is a gain variable which
converts digital full scale levels into dB sound pressure level.

\section{Audio amplification and output\label{sec:Audio-amplification-and}}

The gained bands of audio are summed and output
\[
y=\sum_{m=0}^{M-1}g_{m}h_{m}*s
\]
At this point overlap add sums the last block of audio to the current
block of audio to generate the receiver's output audio signal ($r_{n}$)
\[
r_{n}=y_{n-N/2}*w_{n}+y_{n}*w_{n}
\]

\section{The prescription neural network\label{sec:The-presciption-neural}}

The neural network will input signal levels per band for each block
of audio and output the required signal gain per band ($g_{m}$).
All neural network pre and post conditioning are applied in this block
of processing.

The neural network can be multi-layer and have arbitrary non-linear
layer output functions. The implementation of the prescription neural
network \cite{Flax-NL2P-2012} is proprietary software.

\section{Conclusion}

This article replaces traditional hearing aid cores which are based
on compressors (see Figure \ref{fig:A-digital-hearing}) with the
a suitable SII maximising neural network (see Figure \ref{fig:A-neural-network}).
A traditional prescription system such as NAL-NL2 can be placed directly
onto the users hearing aid in the form of a personal prescription
neural network. This personal prescription neural network can then
be trained to learn the user's preference in amplification. With time
as the free parameters in the neural network are increased in number,
more complex features and learning may be accomplished.

A suitable FIR filter for this hearing aid is defined in Section \ref{sec:Log-banded-filter}
which targets a half block input/output delay to allow a roughly 3
ms system latency which matches the optimal operating latency for
the NAL-NL2 prescription algorithm. A simple sound level meter and
amplification strategy is also defined in Sections \ref{sec:The-sound-level}
and \ref{sec:Audio-amplification-and}. 

\bibliographystyle{plain}
\bibliography{3_home_flatmax_flatmax_personal_work_research_HearingAid_bib}

\end{document}